\documentclass[12pt]{article}
\title{Impact of CP Violation on $|V_{e3}|$}
\author{G. Couture, C. Hamzaoui and A.M. Lussier \\
D\'epartement des Sciences de la Terre et de l'Atmosph\`ere, \\
Universit\'e du Qu\'ebec \`a Montr\'eal\\
C.P. 8888, Succ. Centre-Ville\\
Montr\'eal, H3C 3P8}
\date{}
\begin{document}
\maketitle
\thispagestyle{empty}
\begin{abstract}
We study the impact of the positivity of the measure of CP violation  
in neutrino oscillations on the values of $|V_{e3}|$ and 
$|V_{\mu 1}|$. 
We use the available information on the solar and atmospheric 
mixing parameters $|V_{e2}|$ and $|V_{\mu3}|$ provided by neutrino oscillation 
data.
We conclude that large values of $|V_{e3}|$ are preferred at 95$\%$ 
confidence level. This, in turn, favours large values for $J_{lepton}$ 
compared to CP violation in the quark sector.
\end{abstract}
\newpage
\section{Introduction}
There are now compelling observations that neutrinos (leptons) do mix like 
quarks and have non zero masses \cite{fukuda,SNO,kamland,k2k,chooz}. 
Therefore we have a mixing 
matrix, analogous to that of the quark sector. In the 
absence of a fundamental theory of the origin of mass, this matrix is 
parametrised by several indenpendant parameters. 
Clearly, one of the important goals of 
particle physics in the next years will be to determine these parameters 
with the highest precision possible. One hopes that the knowledge of this 
matrix will point the way to a more fundamental theory of lepton mixings. 
Lepton mixing also has far reaching consequences in leptogenesis \cite{lepto} 
and in baryogenesis \cite{baryo}.

Over the last few years, several groups have strived to set bounds on the 
mixing parameters of the leptonic sector. 
Two preferred scenarios have emerged: the bi-maximal scenario \cite{bimax} and 
the tri-bimaximal scenario \cite{tri}. All these scenarios are summarized in 
reference \cite{xing}. 
Recent data on neutrinos oscillations \cite{pdg} favour 
the tri-bimaximal scenario of lepton mixing.

The purpose of this article is to study the role played by the measure of CP 
violation \cite{me4}, namely $J$, in the structure of the 
leptonic mixing matrix \cite{PMNS}. A similar study has proven usefull before 
in the quark sector \cite{hamzaoui1}. We will show that by 
using the positivity of $J^2$, it is possible to restrict the range of 
$|V_{e3}|^2$ and $|V_{\mu 1}|^2$ in a usefull way. 
As we will see shortly, the maximum amount of CP violation in neutrino 
oscillations is gauged by the value of $\sin \theta_{13}$ \cite{b}, 
which itself is closely related 
to $|V_{e3}|$. We will then concentrate on this last parameter and show that,
at 95$\%$ confidence level (c.l.),  $0.0065 \leq |V_{e3}|^2 \leq 0.05$.

In the next section, we will describe the parametrization that we used for our 
analysis, which we will explain in the third section. In the fourth 
and fifth sections we will present our results on $|V_{e3}|^2$ and 
$|V_{\mu 1}|^2$, and on $J^2$. We will conclude in the last section.

\section{Rephasing Invariant Parametrization of the Mixing Matrix}

We will assume that we have three neutrino flavors. In this case, 
the elements of the lepton mixing matrix are generally defined by

\begin{eqnarray}
\pmatrix{ \nu_e \cr
\nu_{\mu} \cr
\nu_{\tau} } = \pmatrix{V_{e1} & V_{e2} & V_{e3} \cr
V_{\mu 1} & V_{\mu 2} & V_{\mu 3} \cr
V_{\tau 1} & V_{\tau 2}  & V_{\tau 3} \cr}
\pmatrix{ \nu_1 \cr
\nu_2 \cr
\nu_3 }
\end{eqnarray}

\noindent
In terms of mixing angles, the mixing matrix becomes \cite{pdg} 
\begin{eqnarray}
V_{PMNS} = U P=\pmatrix{ c_{12}s_{13} & s_{12}c_{13} & s_{13}e^{-i\delta} \cr
-s_{12}c_{23}-c_{12}s_{23}s_{13}e^{i\delta} & 
c_{12}c_{23}-s_{12}s_{23}s_{13}e^{i\delta} & s_{23}c_{13} \cr
s_{12}s_{23}-c_{12}c_{23}s_{13}e^{i\delta} & 
c_{12}s_{23}-c_{12}c_{23}s_{13}e^{i\delta} & c_{23}c_{13} } P
\end{eqnarray}

\noindent
where $U$ is a Cabibbo-Kobayashi-Maskawa like unitary matrix and CP-violation 
{\it \`a la Dirac} is implemented through the phase $\delta$. 
P is a matrix that contains 
Majorana CP violating phases. These phases are not relevant to our 
analysis as long as we consider only the data from oscillation experiments  
\cite{branco}. Therefore, we can ignore them in what follows.

In order to describe CP-violation in neutrino oscillations, 
the mixing matrix $V_{PMNS}$ can be parametrized in terms of four moduli
\cite{cherif}; we choose the following four independant moduli: 
$|V_{e2}|$, $|V_{e3}|$, $|V_{\mu1}|$ and $|V_{\mu 3}|$.

As the $V_{ij}$ are complex quantities, setting $V_{e3} = 0$ gives us 
two constraints: 
$|V_{e3}|=0$ and $|V_{\mu1}|= |V_{e2}|\sqrt{1-|V_{\mu 3}|^2}$.
The whole mixing matrix is then 
described by $|V_{e2}|$ and $|V_{\mu 3}|$. 
Consequently, we have:

\begin{eqnarray}
V_{PMNS} = \pmatrix{\sqrt{1-|V_{e2}|^2} & |V_{e2}| & 0 \cr
-|V_{e2}|\sqrt{1-|V_{\mu 3}|^2} & \sqrt{(1-|V_{e2}|^2)(1-|V_{\mu 3}|^2)} & 
|V_{\mu 3}| \cr
|V_{e2}| |V_{\mu 3}| & -|V_{\mu 3}| \sqrt{1-|V_{e2}|^2} & \sqrt{1-|V_{\mu 3}|^2} 
\cr} 
\end{eqnarray}

It is now possible to estimate the maximal value for $|V_{\mu1}|$. 
In the bi-maximal scenario, 
one takes $|V_{\mu3}|=|V_{e2}|=\frac{1}{\sqrt{2}}$ which leads to 
$|V_{\mu1}|_{bi-max}=\frac{1}{2}$ 
\noindent and 
\begin{equation}
U_{bi-max}= \left (
\begin{array}{ccc}
 \frac{1}{\sqrt{2}} & \frac{1}{\sqrt{2}} & 0 \\
-\frac{1}{2} & \frac{1}{2} & \frac{1}{\sqrt{2}} \\
\frac{1}{2} & -\frac{1}{2} & \frac{1}{\sqrt{2}}
\end{array} \right )
\end{equation} 

In the tri-bimax scenario, one sets 
$|V_{e2}|=\frac{1}{\sqrt{3}}$ and $|V_{\mu3}|=\frac{1}{\sqrt{2}}$. This 
leads to $|V_{\mu1}|_{tri-bimax}=\frac{1}{\sqrt{6}}$ 
\noindent and 

\begin{equation}
V_{tri-bimax}= \left (
\begin{array}{ccc}
\sqrt{\frac{2}{3}} & \frac{1}{\sqrt{3}} & 0 \\
-\frac{1}{\sqrt{6}} & \frac{1}{\sqrt{3}} & \frac{1}{\sqrt{2}} \\
\frac{1}{\sqrt{6}} & -\frac{1}{\sqrt{3}} & \frac{1}{\sqrt{2}}
\end{array} \right )
\end{equation} 

We argue that this maximal value of $|V_{\mu1}|$ is quite reasonable as any 
non-zero value of $V_{e3}$ will increase it or decrease it by an amount 
close to $|V_{e3}|$ or smaller \cite{hamzaoui1, cherif}.  
Nonetheless, we will study 
two cases later on: $|V_{\mu 1}|^2_{max}=0.25$ and $0.35$.

\noindent
The exact expressions between the moduli of the elements of the mixing matrix 
($V_{PMNS}$) and the mixing angles experimentally accessible, namely
$\theta_{sun}$, $\theta_{atm}$ and $\theta_{13}$, can be found in Xing
\cite{xing}; setting $|V_{e3}| =0$ leads to $|V_{e2}| = sin\theta_{sun}$ and 
$|V_{\mu 3}| = sin\theta_{atm}$. 
The expressions that we need here are:
\begin{eqnarray}
|V_{e3}|^2 & =& sin^2\theta_{13} \\
|V_{e2}|^2 & =& sin^2\theta_{12}~cos^2\theta_{13}\\
|V_{\mu 3}|^2 & =&sin^2\theta_{23}~cos^2\theta_{13}
\end{eqnarray}

Our current knowledge on the magnitude of the elements of the leptonic mixing 
matrix comes from experiments on neutrino oscillation and can be summarized as 
follows\cite{bahcal}:

\begin{eqnarray}
\sin^2 \theta_{12} & = & 0.3\pm0.08 \\
\sin^2 \theta_{23} & = & 0.5\pm0.18 \\
\sin^2 \theta_{13} & \leq & 0.05~~{\rm at}~3\sigma
\end{eqnarray}
\noindent
Note that these values are close to those of the tri-bimaximal scenario.

The measure of CP violation, $J$, is given in general as: 

\begin{eqnarray}
J = \Im m(V_{e2} V_{\mu 3} V^*_{e3} V^*_{\mu 2}) 
\end{eqnarray}\\
\noindent
It has been shown before that if any element of the mixing matrix goes 
to zero, CP violation {\it \`a la Dirac} vanishes \cite{cherif}. 
However, the vanishing of CP 
violation does not imply that any element of the mixing matrix is zero.

\noindent
From the previous expression for $J$, we can infer an upper bound:
\begin{eqnarray}
|J| \leq  |V_{e2}| ~ |V_{\mu 3}| ~ |V_{e3}| ~ |V_{\mu 2}|
\end{eqnarray}\\

\noindent
Using the values of the tri-bimaximal scenario   
($|V_{e2}|= 1/\sqrt{3}=|V_{\mu 2}|,~|V_{\mu 3}|=1/\sqrt{2}$) 
we can get a rough estimate on the maximum value of  
$|J|$:
 
\begin{eqnarray}
|J| \leq \frac{|V_{e3}|}{3\sqrt{2}} 
\end{eqnarray}
\noindent
Numerically, by taking the maximal value of $|V_{e3}|\sim \sqrt{0.05}$, we have 
$|J|_{max}\approx 0.05$

As can be seen from eqs. 1 and 2,  $|V_{e3}|$ has the simplest relation to
$V_{PMNS}$ and CP violation. It is also a measure of the maximum amount of 
CP violation possible in the leptonic sector.  
We will therefore pay special attention to this parameter.

\noindent
In terms of our 4 independant parameters, the measure of CP-violation is 
given by \cite{cherif} 

\begin{eqnarray}
& J^2 = (1-|V_{e3}|^2-|V_{e2}|^2)|V_{e3}|^2|V_{\mu1}|^2|V_{\mu3}|^2\cr
& -\frac{1}{4}\left ( 
|V_{e2}|^2-|V_{\mu1}|^2+|V_{\mu1}|^2|V_{e3}|^2-|V_{\mu3}|^2|V_{e3}|^2-|V_{\mu3}|^2
|V_{e2}|^2 \right )^2
\end{eqnarray}\label{Jcarre}

\section{Analysis}
For our analysis, we use only the known mixing parameters: the solar 
mixing $\theta_{12}$, the atmospheric mixing $\theta_{23}$. We will use 
the positivity of $J^2$ to set an interval on $|V_{e3}|^2$ and 
$|V_{\mu 1}|^2$ consistent with data. 
In the following, we will assume that $cos^2\theta_{13}\approx 1$ and later on 
we will estimate how a variation in this value could affect our results. This 
means that we use $|V_{e2}|^2\approx sin^2\theta_{12}= 0.3\pm 0.08$ and 
$|V_{\mu 3}|^2\approx sin^2\theta_{23}= 0.5\pm 0.18$ and set the maximal value 
allowed for $|V_{\mu1}|$ at that obtained from the bi-maximal scenario:

\begin{equation}
|V_{\mu 1}|_{max} = 0.5
\end{equation}

We performed our numerical analysis as follows. Using the central values 
for $|V_{e2}|^2$ and $|V_{\mu 3}|^2$ we use eq.-15 and require that $J^2$ be 
positive. This gives us a range of values for $|V_{e3}|^2$ and $|V_{\mu 1}|^2$,
as can be seen on fig. 1: any combination within the parabola leads to a 
positive value of $J^2$. In order to take into account the uncertainties in 
$|V_{e2}|^2$ and $|V_{\mu 3}|^2$, we used a Monte Carlo method. We generated 
values for $|V_{e2}|^2$ and for $|V_{\mu 3}|^2$, independantly, according to 
the Gaussian distribution $exp[-0.5((x-x_c)/\sigma_x)^2]$ centered at $x_c$ 
and with a standard deviation of $\sigma_x$. 
We covered the whole range from 0 to 0.6 for $|V_{e2}|^2$ and  0 to 1 
for $|V_{\mu 3}|^2$ and verified that our distributions were centered at 0.3 
and 0.5 with standard deviations of 0.08 and 0.18. For each combination of 
$|V_{e2}|^2$ and $|V_{\mu 3}|^2$, we compute a curve similar to that of fig. 1.
The numbers that we present here were obtained from 50,000 such 
combinations of $(|V_{e2}|^2,|V_{\mu 3}|^2)$. Once we have these 50,000 
curves, we scan the $|V_{e3}|^2-|V_{\mu 1}|^2$ plane and 
count how many combinations of $(|V_{e2}|^2,|V_{\mu 3}|^2)$  allow a given 
point on the grid; we used a grid of 40,000 points (200 on each axis). If a
point is allowed by the lower branch of a parabola, it gains a weight of 
1, similarly for the upper branch of a parabola. If a point of the grid is 
rejected by a lower branch of a parabola, it looses a weight of 1; similarly 
for the upper branch of a parabola. This results in a given point of the grid 
gaining a weight of 2 if it is allowed by a parabola and gaining a weight of 0 
if it is not allowed. This procedure leads  
to fig. 2-a where the probability of a given combination is given by the volume 
under the surface. The total volume under the surface is unity since we know 
that the final combination of parameters will be somewhere in our parameter 
space. The shape of this figure is easily understood: the base of the parabola 
will move a lot along the $|V_{\mu 1}|^2$ axis as values of $|V_{e2}|^2$ and 
$|V_{\mu 3}|^2$ are varied, thus excluding several points in the 
$|V_{e3}|^2-|V_{\mu 1}|^2$ plane that are close to the $|V_{\mu 1}|^2$ axis,
while points for larger values of $|V_{e3}|^2$ will be accepted by several 
parabolas. In order to put a confidence level on the combination 
$(|V_{e3}|^2,|V_{\mu 1}|^2)$, one simply slices the surface of fig. 2-a  
horizontally, starting from the top. Each slicing intersects the surface 
and the volume under the surface defined by this intersection represents the 
probability of finding the combinations $(|V_{e3}|^2,|V_{\mu 1}|^2)$. 
When this volume reaches 68$\%$ of the total volume, 
the curve that results of the 
intersection of the surface and the slicing is the 68$\%$ c.l. 
curve in the $|V_{e3}|^2$ and $|V_{\mu 1}|^2$ parameter space. The same
applies for 
the 90$\%$ and 95$\%$ c.l. curves. This leads to figures 2-b.

In order to set c.l. limits on 
$|V_{e3}|^2$ alone, one has to integrate over $|V_{\mu 1}|^2$. 
Starting with fig. 2-a, it is easy to integrate over 
$|V_{\mu 1}|^2$: we simply add all the volume elements along an axis. This 
results in fig. 3, where the total surface under the curve is 1 since the 
physical, final parameter must be under this curve. In order to establish 
our c.l.s, we proceed as before and slice 
our figure horizontally, starting at the top. The intersection of our curve 
and the slicing, will define a certain range for $|V_{e3}|^2$ and when this 
range will cover a surface under the curve that represents 68$\%$ of the 
total surface, this range will be our 68$\%$ c.l. for 
$|V_{e3}|^2$. The same procedure applies for the 90$\%$ and 95$\%$ confidence 
level ranges. 

\section{Results}
According to this procedure, we can say that the values of  
$|V_{e2}|^2= 0.3 \pm 0.08$ and $|V_{\mu 3}|^2 = 0.5\pm 0.18$ and the allowed 
ranges $0\leq |V_{e3}|^2\leq 0.05$ and  $0\leq |V_{\mu 1}|^2 \leq 0.25$
lead to a 68$\%$, a 90$\%$, and a 95$\%$ c.l. ranges in the 
$|V_{e3}|^2-|V_{\mu 1}|^2$ parameter space as shown on figure 2-b. 
Globally, $|V_{\mu 1}|^2$ has to be larger than 0.027 at 95$\%$ 
c.l. and $|V_{e3}|^2$ has to be larger than 0.0017, also at 
95$\%$ c.l.. These bounds become 0.034 and 0.0033, respectively, 
at 90$\%$ c.l.. 
If we intergrate over $|V_{\mu 1}|^2$, we can say the 68$\%$ 
c.l. range for $|V_{e3}|^2$ is 0.0228 -- 0.05; 
the  90$\%$ c.l. range is  
0.0103 -- 0.05 and the 95$\%$ c.l. range is 0.0065 -- 0.05. 
Therefore, we can say that values of $|V_{e3}|^2$ smaller than 0.0065 are 
excluded at the 95$\%$ c.l..

We can also project ourselves in the future and ask what these limits on 
$|V_{e3}|^2$, after integration over $|V_{\mu 1}|^2$, will 
become once the errors on $|V_{e2}|^2$ and $ |V_{\mu 3}|^2$ will have been 
reduced. We performed the same analysis as above but we reduced the error on 
$|V_{e2}|^2$ and $ |V_{\mu 3}|^2$ to 0.02 and 0.04, respectively. 
We studied 9 different combination of 
$|V_{e2}|^2$ and $ |V_{\mu 3}|^2$: $(0.22,~0.30,~0.38)~\times ~ 
(0.32,~0.50,~0.68)$, respectively. The {\it bump} that appears on fig. 2-a 
moves substantially over the allowed ranges of $|V_{e3}|^2$ and 
$|V_{\mu 1}|^2$ but once we integrate over $|V_{\mu 1}|^2$ the allowed ranges 
of $|V_{e3}|^2$ change very little with the different combinations: 
0.023 -- 0.05 at 68$\%$ c.l., 0.011 -- 0.05 at 90$\%$ c.l. and 0.0068 -- 0.05 
at 95$\%$ c.l. The only exception being (0.38, 0.50) where they are 
0.021 -- 0.05, 0.0093 -- 0.05, and 0.0058 -- 0.05.

One could ask how these bounds on $|V_{e3}|^2$ will vary if we allow the ranges
on $|V_{e3}|^2$ and $|V_{\mu 1}|^2$ to vary. We explored this possibility and 
changed the ranges to $0\leq |V_{e3}|^2\leq 0.07$ and 
$0\leq |V_{\mu 1}|^2 \leq 0.35$. From fig. 2-a, we can expect the general 
behaviour to continue for larger values of $|V_{e3}|^2$ and $|V_{\mu 1}|^2$. 
The different c.l. ranges obtained from the central experimental
values of $|V_{e2}|^2$ and $|V_{\mu 3}|^2$ with their current errors, are 
shown on fig. 2-c. The different ranges are: 0.010 -- 0.07 on $|V_{e3}|^2$ and 
0.051 -- 0.285 on 
$|V_{\mu 1}|^2$ at 68$\%$ c.l., 0.0025 -- 0.07 on $|V_{e3}|^2$ 
and 0.021 -- 0.333 on 
$|V_{\mu 1}|^2$ at 90$\%$ c.l., 0.0011 -- 0.07 on $|V_{e3}|^2$ 
and 0.014 -- 0.35 on 
$|V_{\mu 1}|^2$ at 95$\%$. Furthermore, when we integrate over $|V_{\mu 1}|^2$,
the c.l. ranges on $|V_{e3}|^2$ are 0.0322 -- 0.07, 0.0147 -- 0.07, 
and 0.0091 -- 0.07, at 68$\%$, 90$\%$, and 95$\%$, respectively. This 
strenghtens our statement that values of $|V_{e3}|^2$ smaller than 0.0065 are
excluded at the 95$\%$ c.l..

In this context of larger bounds on $|V_{e3}|^2$ and $|V_{\mu 1}|^2$, we also 
projected ourselves in the future and redid the study with smaller errors on 
$|V_{e2}|^2$ and $ |V_{\mu 3}|^2$, namely 0.02 and 0.04. The results are that, 
again the different c.l. limits are rather insensitive to the different
combinations of $|V_{e2}|^2$ and $ |V_{\mu 3}|^2$ used and read as follows: 
0.033 -- 0.07 at 68$\%$ c.l., 0.015 -- 0.07 at 90$\%$ c.l. and 0.0095 -- 0.07
at 95$\%$ c.l. The only exception being (0.38, 0.32) where they are
0.030 -- 0.07, 0.014 -- 0.07, and 0.0088 -- 0.07. 

\section{What can we say about $J^2$ ?}
At this point, one can ask what the typical values of $J^2$ in the allowed 
ranges are. On figure 4, we simply plot $J^2$ as a function of $|V_{e3}|^2$ and 
$|V_{\mu 1}|^2$ for $|V_{e2}|^2= 0.3 $ and $|V_{\mu 3}|^2 = 0.5$; the 
intersection of the {\it cone} with the plane is our fig. 1. On this figure, 
the maximum value of $J^2$ is $2.42\times 10^{-3}$. One also sees that $J^2$  
decreases rapidly as we move away from the peak. Note however that for 
the allowed ranges discussed previously, the values taken by $J^2$ will be 
typically larger than $3\times 10^{-4}$. In order to give a better estimate, 
we generated combinations of our 4 parameters with the weights given by
figure 2-a and calculated $J^2$ for each combination. The results are that for 
all our c.l. ranges, $J^2$ takes values between 
 $\sim 3\times 10^{-8}$ 
and $2.7\times 10^{-3}$ while the standard deviations of the 
distributions are constant at $0.6\times 10^{-3}$. The average values of the 
distributions vary a little: $1.1\times 10^{-3}$ over the 68$\%$ confidence 
level range, $0.98\times 10^{-3}$ over the 90$\%$, and 
$0.97\times 10^{-3}$ over the 95$\%$ cl range. Globally, these distributions 
do not vary much over our different c.l. ranges. Comparing figures 2-b-c 
with figure 1, it is clear that we cannot exclude $J^2 = 0$ since the
boundaries between $J^2$ positive and negative are well within our confidence 
level ranges for many combinations of $|V_{e2}|^2$ and $|V_{\mu 3}|^2$.
Our real minimal value is in fact zero and only numerical 
accuracy made it non-zero. What is important here is not the minimal 
value but rather  
the average values over the different domains: these are clearly far from zero 
and we can say that smaller values of $J^2$ are disfavoured.

As for $J$, the results are practically independant of the c.l. 
ranges and 
read as $J_{min} \sim 2 \times 10^{-4}$, $J_{max}= 0.05$, with an average 
value of 0.03 and a standard deviation of 0.01. The average values are 
the more meaningful results and small values are again strongly disfavoured. 
These are the strongest statements that we can make about $J^2$ and $J$.

From the previous variations on our parameters, it is clear that our initial 
approximation ({\it ie} $|V_{e2}|^2\approx sin^2\theta_{12}= 0.3\pm 0.08$ and
$|V_{\mu 3}|^2\approx sin^2\theta_{23}= 0.5\pm 0.18$) has very little impact 
on our results.

\section{Conclusion}

 The positivity of $J^2$ has proven a useful tool in setting constraints on 
different parameters of the leptonic mixing sector. Using only this variable, 
we can say that values of $|V_{e3}|^2$ smaller than $0.0065$ are excluded at 
95$\%$, stronly disfavoured, by current data on $|V_{e2}|^2$ and 
$|V_{\mu 3}|^2$. 
Such large values are in fact preferred by some authors as they can help in 
establishing the complementarity
relations between quark and lepton mixing \cite{mohapatra}. From fig. 4, one
sees that a large value of $|V_{e3}|^2$ is not a guarantee of CP violation
but simply favours large CP violation.

We can also say that $J^2$ must be smaller than 
$2.7\times 10^{-3}$ and that values much 
smaller than $1\times 10^{-3}$ are strongly disfavoured. This translates 
into $J_{max}\approx 0.052$ and values much smaller than 0.03 being also 
strongly disfavoured. Such values of $J_{lepton}$ are typically 1000 times 
larger than values of $J_{quark}$.

When looking at figures 1 and 2, one can see that the positivity of $J^2$ will
not be able to exclude $|V_{e3}|^2 = 0$ since the parabola will always
extend all the way down to $|V_{e3}|^2 = 0$, no matter what the experimental
precision on $|V_{e2}|^2$ and $|V_{\mu 3}|^2$ is. All we can do is set some
confidence limits.

Considering figures 1 and 4, it is clear that the tightest indirect contraints 
on $|V_{e3}|^2$ and $|V_{\mu 1}|^2$ will come from an upper limit or a direct 
measurement of CP violation in neutrino oscillations; from a limit or a direct 
measurement of $J_{lepton}$. This would reduce the space to a narrow strip 
following the parabola of fig. 1. Short of such a measurement, one will have to 
rely on independant measurement of $|V_{e3}|^2$ and $|V_{\mu 1}|^2$.  
A measurement of $|V_{\mu 1}|^2$ {\it far
away} from 0.15 (assuming that $|V_{e2}|^2$ and $|V_{\mu 3}|^2$ keep their
current central values) would lead to the smallest possible range on
$|V_{e3}|^2$ and push it far from 0, thereby increasing the probability of 
large CP violation. Similarly, a measurement of $|V_{e3}|^2$ with a very small 
value would put the tightest limits on $|V_{\mu 1}|^2$ and at the same time 
would reduce greatly the possibility of large value of $|J|^2$. 
Assuming that a small value
of $|V_{e3}|^2$ will be difficult to measure, one concludes that experimental
efforts should be devoted to measuring $|V_{\mu 1}|^2$ since this could
potentially strongly restrict the range of $|V_{e3}|^2$. 
Let's not forget that 
complete determination of the neutrino mixing matrix will shed some light on
the deep structure of the leptonic spectrum and help us understand better
the properties of neutrinos and mass generation in general.

\section{Acknowldgements}
\noindent
It is a pleasure to thank Dr. G. Azuelos for suggesting the use of Monte Carlo
methods. This work was supported in part by NSERC of Canada.

\section{Figures Captions}
\noindent
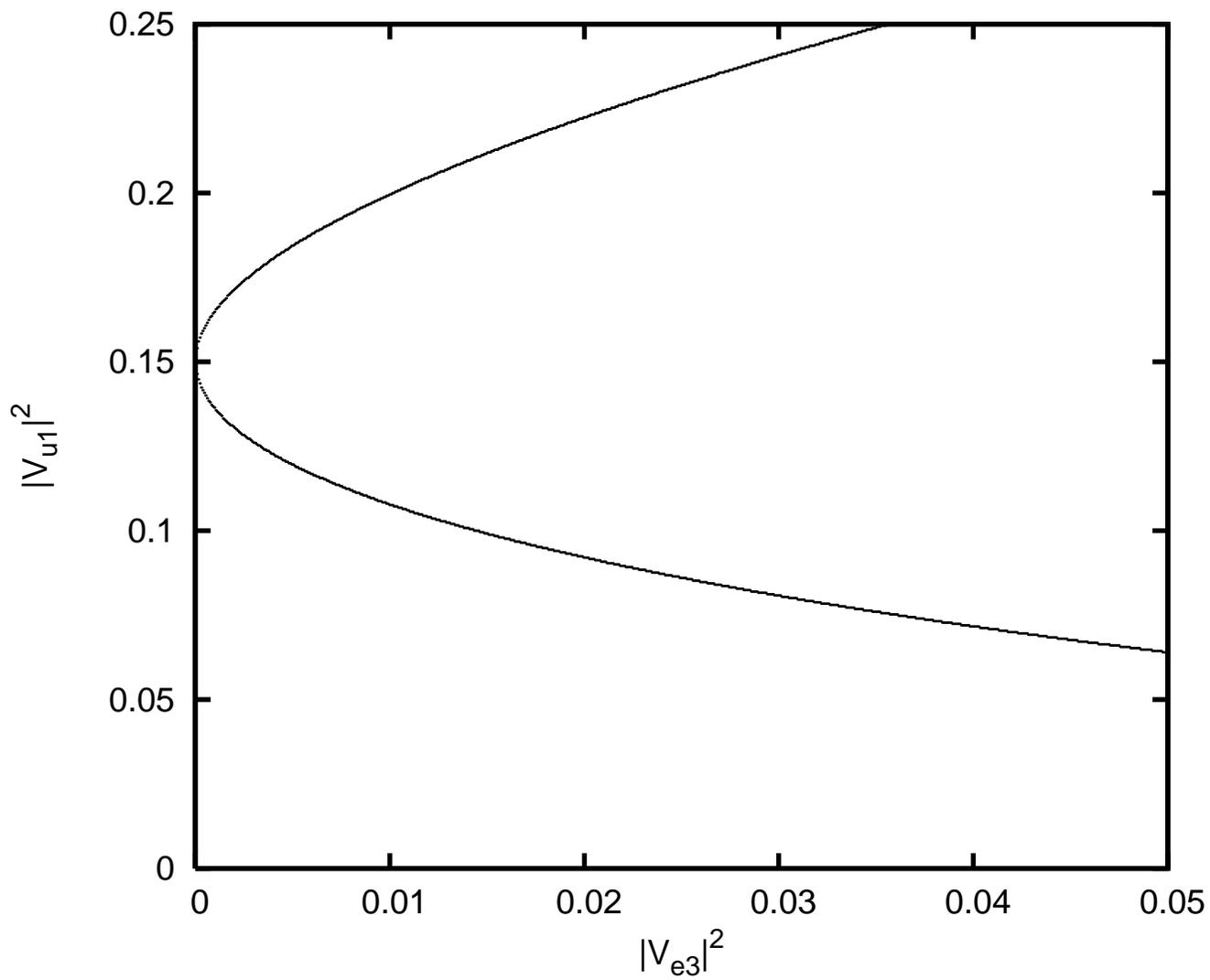
Figure 1: Limits in the $|V_{e3}|^2-|V_{\mu 1}|^2$ plane obtained from the 
positivity of $J^2$. The parabola represents $J^2 = 0$; the space 
inside represents positive values of $J^2$. 
\hfil\hfil\break
\noindent
Figure 2-a: Probability distribution in the $|V_{e3}|^2-|V_{\mu 1}|^2$ plane 
obtained from 50,000 curves similar to those of figure 1. The vertical axis 
is in arbitrary units as the volume under the curve, representing the total 
probability, is unity.
\hfil\hfil\break
\noindent
Figure 2-b: Confidence levels obtained from figure 2-a. The 68$\%$ c.l. has the 
smallest surface while the 95$\%$ c.l. has the largest.
\hfil\hfil\break
\noindent
Figure 2-c: Similar to figure 2-b but with $0\leq |V_{e3}|^2\leq 0.07$ and 
$0\leq |V_{\mu 1}|^2\leq 0.35$.
\hfil\hfil\break
\noindent
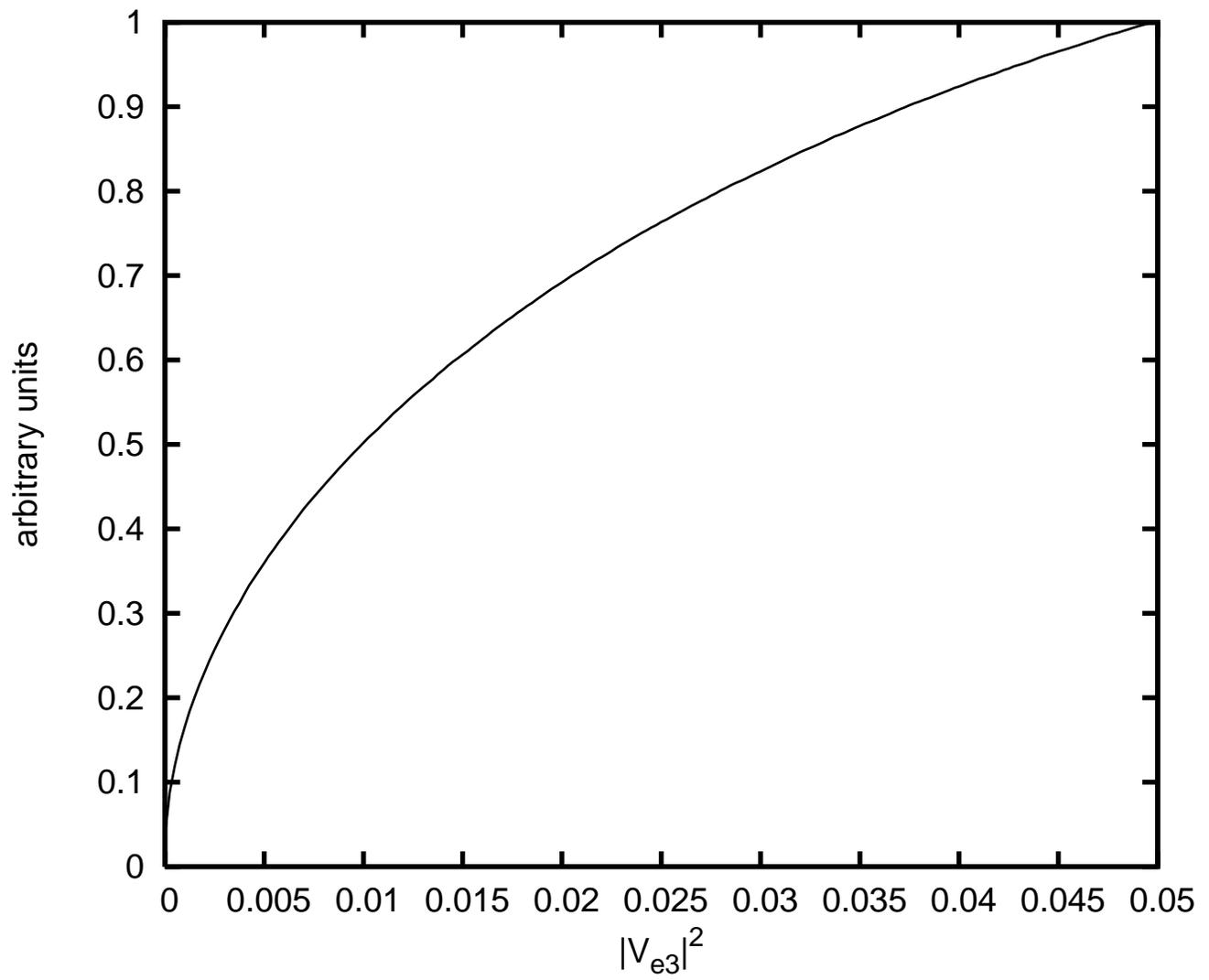
Figure 3: Probability distribution of $|V_{e3}|^2$ after integrating the 
$|V_{\mu 1}|^2$ axis of figure 2-a. 
\hfil\hfil\break
\noindent
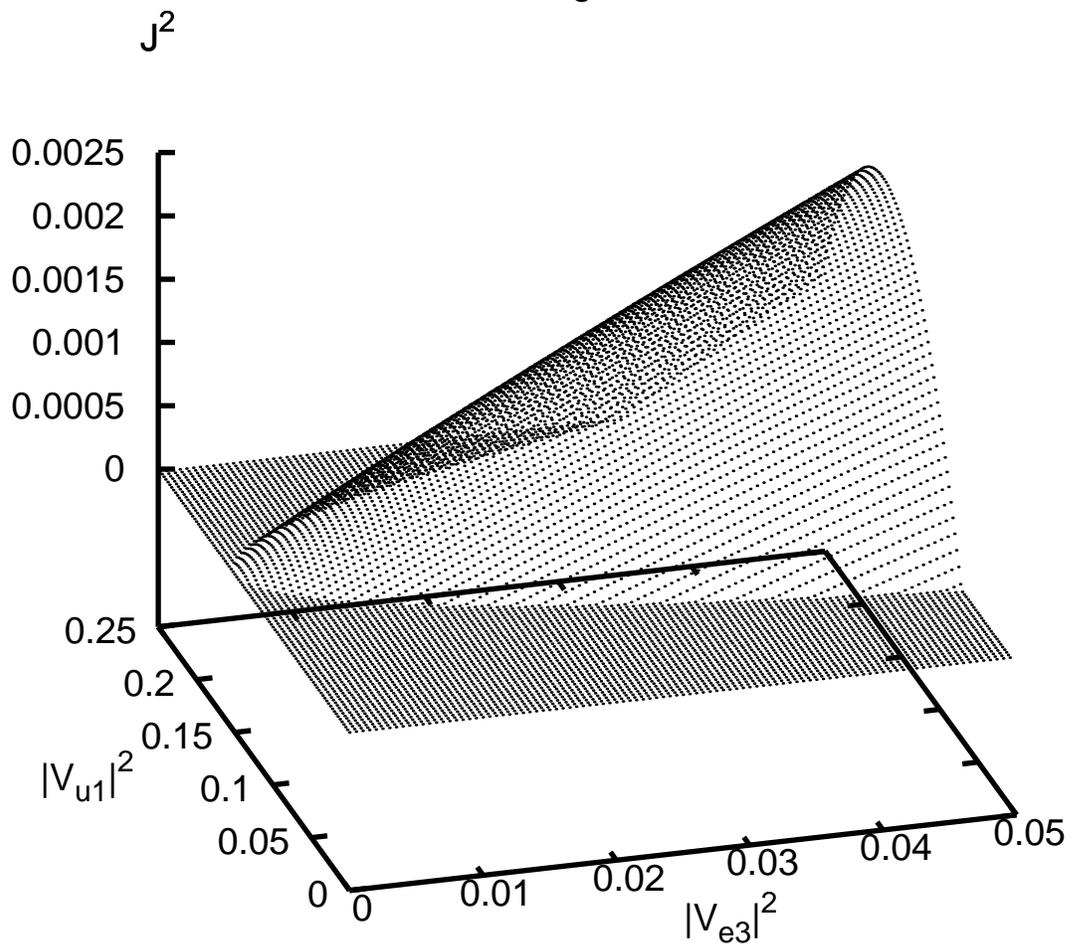
Figure 4: $J^2$ for different values of $|V_{e3}|^2$ and $|V_{\mu 1}|^2$ using 
$|V_{e2}|^2 = 0.3$ and $|V_{\mu 3}|^2 = 0.5$

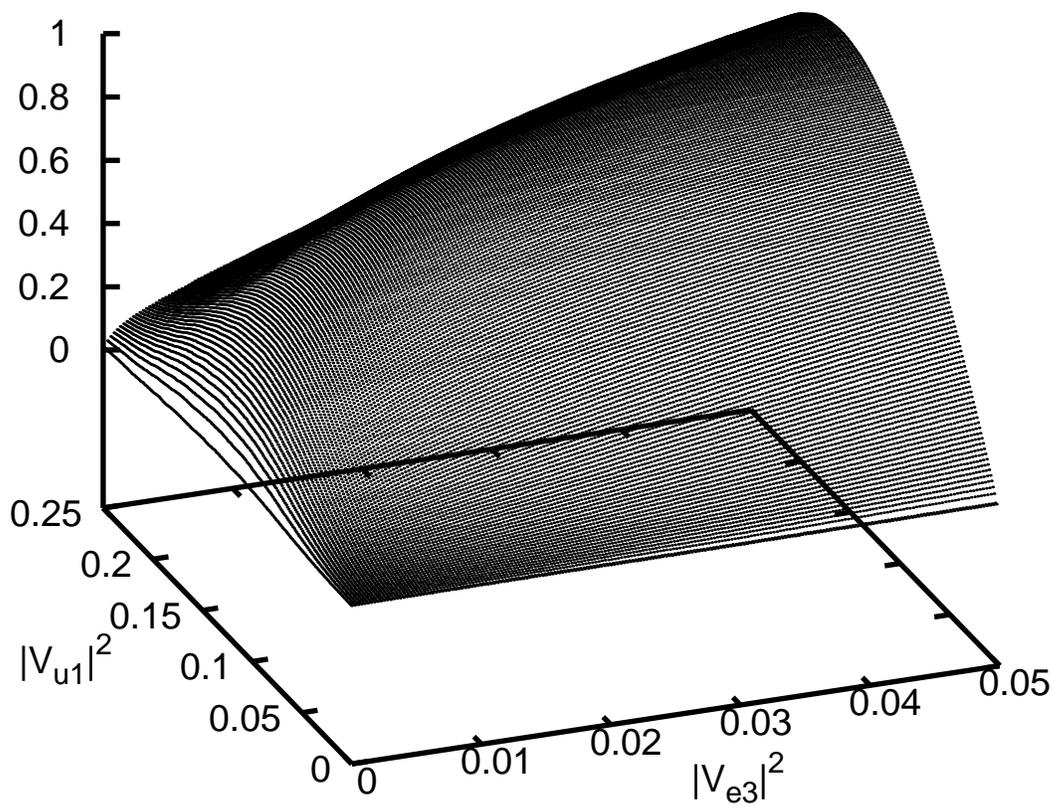
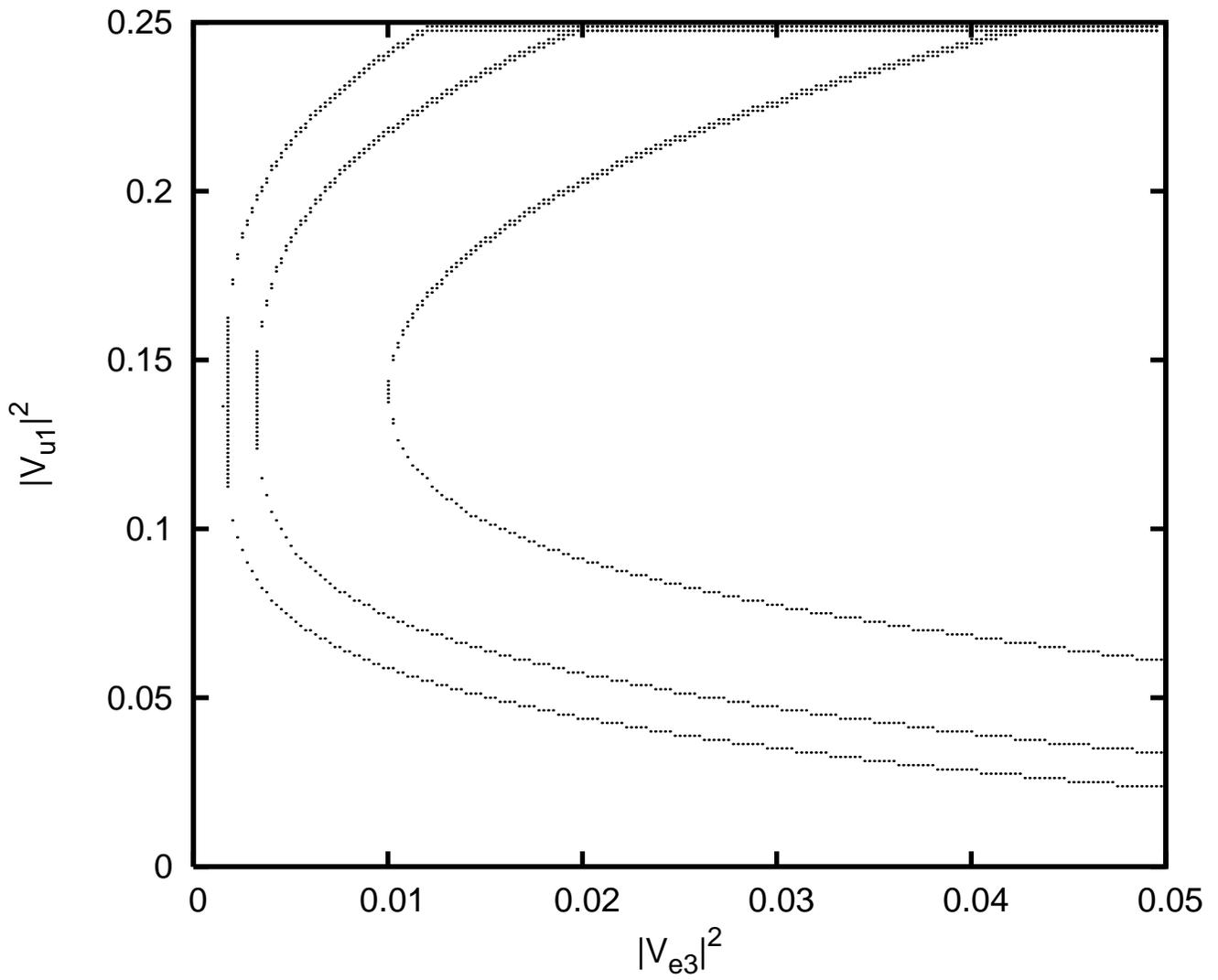
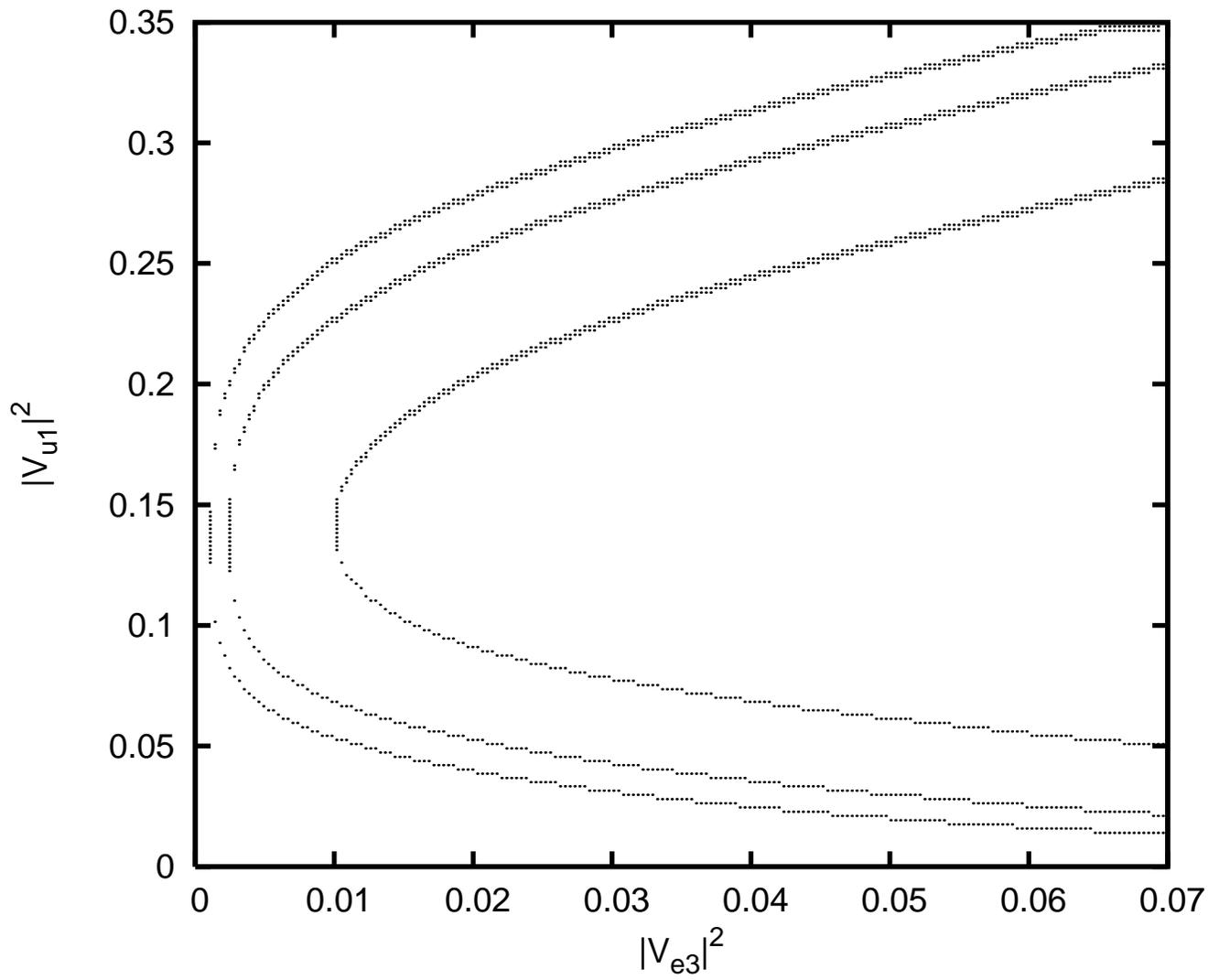
\end{document}